\documentclass[preprint,superscriptaddress]{revtex4-2} 
\usepackage{graphicx} 
\usepackage{amsmath, cases} 
\usepackage[version=3]{mhchem} 
\usepackage{longtable} 

\usepackage{mathtools} 
\usepackage{url} 
\usepackage{verbatim} 
 
\usepackage{epstopdf} 
\usepackage{ upgreek }

\usepackage{hyperref} 
\usepackage{cleveref} 

\usepackage[normalem]{ulem} 
\usepackage{xcolor}

\begin{document} 

\title{Nonempirical Determination of the Optimal TAO-DFT Fictitious Temperature for Molecular Systems in the Lowest Singlet State} 

\author{Bing-Ying Li} 
\affiliation{Department of Physics, National Taiwan University, Taipei 10617, Taiwan} 

\author{Jeng-Da Chai} 
\email[Author to whom correspondence should be addressed. Electronic mail: ]{jdchai@phys.ntu.edu.tw} 
\affiliation{Department of Physics, National Taiwan University, Taipei 10617, Taiwan} 
\affiliation{Center for Theoretical Physics and Center for Quantum Science and Engineering, National Taiwan University, Taipei 10617, Taiwan} 
\affiliation{Physics Division, National Center for Theoretical Sciences, Taipei 10617, Taiwan} 

\date{\today}

\begin{abstract} 

Thermally-assisted-occupation density functional theory (TAO-DFT) [J.-D. Chai, {\em J. Chem. Phys.}, 2012, {\bf 136}, 154104] has been developed to investigate the ground-state properties of large 
electronic systems exhibiting varying multi-reference (MR) character. As a central parameter in TAO-DFT, the fictitious temperature $\theta$ is closely linked to the degree of static correlation in an electronic 
system. While the conventional system-independent $\theta$-scheme is computationally efficient and broadly applicable, a fixed, preassigned $\theta$ can be overly large for single-reference (SR) systems 
yet too small for strongly MR systems. To overcome this limitation, we propose a nonempirical scheme based on spin-symmetry restoration to determine the optimal system-dependent fictitious temperature 
$\theta^{*}$ for molecular systems in their lowest singlet states. Across benchmark tests encompassing both MR systems (the bond dissociation of H$_{2}$, N$_{2}$, Li$_{2}$, and LiH, as well as the 
torsional rotation of twisted ethylene) and pure SR systems (the dissociation of He$_{2}$ and Ne$_{2}$), our nonempirical approach demonstrates superior overall accuracy compared with the conventional 
fixed-$\theta$ scheme. Furthermore, relative to standard Kohn-Sham density functional theory (TAO-DFT with $\theta = 0$), TAO-DFT employing $\theta^{*}$ yields substantial improvements for MR systems 
while preserving comparable accuracy in SR systems. 

\end{abstract} 

\maketitle 

\clearpage

\section{Introduction} 

The fundamental physical principles governing electronic systems are well established; however, their exact application to large systems remains computationally prohibitive \cite{dirac1929}. Exact analytical solutions 
to the Schr\"odinger equation exist only for a limited number of simple systems, necessitating approximate methods for more complex cases \cite{sakurai,griffiths}. Consequently, developing efficient and accurate 
electronic structure methods for general systems remains a central challenge. 

Since its inception in the mid-1960s, Kohn-Sham density functional theory (KS-DFT) \cite{kohn_self-consistent_1965} has become one of the most widely used approaches for predicting the ground-state (GS) 
properties of electronic systems at zero electronic temperature ($\theta_{el} = 0$), owing to its favorable balance between computational efficiency and 
accuracy \cite{RevModPhys.80.3,parr,DFT-exchange,cohen_insights_2008,Yang12}. Nevertheless, because the exact exchange-correlation (xc) energy functional—the core component of KS-DFT—remains unknown, 
practical calculations must rely on approximate xc functionals, which can yield unreliable results in certain regimes \cite{RevModPhys.80.3,parr,DFT-exchange,cohen_insights_2008,Yang12}. 

In particular, accurately predicting the GS properties of multi-reference (MR) systems (i.e., systems exhibiting strong static correlation) poses a major challenge for KS-DFT. For MR systems, conventional local density 
approximation (LDA) \cite{LDA-1,LDA-2}, generalized gradient approximation (GGA) \cite{BLYP-1,BLYP-2,PBE}, global hybrid (GH) \cite{BHHLYP,B3LYP-1,PBE0-1}, range-separated hybrid 
(RSH) \cite{rsh-1,erfgau-1,rsh-2,wb97xd,wb97xd3,wb97m-v,rsh-9}, and double-hybrid (DH) \cite{DH-3,DH-2,DH-4,DH-1,DH-16,DH-9,DH-12} xc functionals often suffer from severe static correlation error 
(SCE) \cite{cohen_insights_2008,Yang12,DFT-exchange}. More fundamentally, even with the exact xc functional, the true GS density of an MR system may not be representable within 
KS-DFT \cite{baerends,KS-rep-1,KS-rep-2,KS-rep-3}. 

To resolve both the SCE and GS density representability issues with minimal computational overhead, thermally-assisted-occupation density functional theory (TAO-DFT) \cite{chai_density_2012} was developed to 
study the GS properties of large electronic systems with varying MR character. In TAO-DFT, the GS electron density is constructed from thermally-assisted-occupation (TAO) orbitals and TAO orbital occupation 
numbers (TOONs), which are governed by the Fermi-Dirac (FD) distribution function at a fictitious temperature $\theta$—the temperature of the non-interacting reference system (commonly termed the TAO reference 
system). With an appropriately chosen $\theta$, the resulting TOONs closely approximate the exact natural orbital occupation numbers (NOONs) \cite{NO}. This significantly improves GS density representability and 
enables the static correlation energy to be accurately captured by the entropy contribution (see, e.g., Section III.E of Chai \cite{chai_density_2012}). Accordingly, $\theta$ should be vanishingly small for 
single-reference (SR) systems, whereas it is strongly system-dependent for MR systems. Notably, TAO-DFT reduces directly to standard KS-DFT when $\theta = 0$, where the entropy contribution vanishes. 

Although the exact exchange-correlation-$\theta$ (xc$\theta$) energy functional (a combined xc and $\theta$-dependent energy functional) in TAO-DFT is also unknown, existing LDA \cite{chai_density_2012}, 
GGA \cite{chai_thermally-assisted-occupation_2014}, GH \cite{chai_role_2017}, and RSH \cite{chai_role_2017,Xuan2019,li_optimization_2025} xc$\theta$ functionals can be readily applied to large MR 
systems \cite{chai_density_2012,chai_thermally-assisted-occupation_2014,chai_role_2017,li_optimization_2025,wu_electronic_2015,yeh_role_2016,seenithurai_electronic_2019,seenithurai_tao-dft_2020}. 
Furthermore, several extensions of TAO-DFT \cite{li_optimization_2025,tao-aimd,RT-TAO,tao-pcm,ft-tao} have been developed recently to broaden its applicability. 

In TAO-DFT, the fictitious temperature $\theta$ controls the TOON distribution and directly governs how much static correlation is captured by the entropy contribution \cite{chai_density_2012}. Existing schemes for 
determining $\theta$ fall into two main categories: system-independent \cite{chai_density_2012,chai_thermally-assisted-occupation_2014,chai_role_2017,li_optimization_2025,chen_tao-dft_2022} and 
system-dependent \cite{lin_self-consistent_2017} schemes. System-independent $\theta$-schemes assign a fixed $\theta$ value to a given xc$\theta$ functional regardless of the system under study. For example, 
Chen and Chai \cite{chen_tao-dft_2022} developed a nonempirical approach based on a reference electronic system (RES), which defines a theoretical boundary state lying between the SR and MR regimes; the 
optimal $\theta$ is then derived directly from the energy gap and orbital occupation numbers of this RES. In particular, this RES model was adopted to determine the optimal system-independent $\theta$ values for 
the LDA, GGA, GH, and RSH xc$\theta$ functionals in TAO-DFT \cite{chen_tao-dft_2022,li_optimization_2025}. 

While system-independent $\theta$-schemes are computationally efficient and reasonably accurate across a broad range of systems, a preassigned fixed $\theta$ can occasionally over-correlate SR systems (when 
$\theta$ is too large) or under-correlate strongly MR systems (when $\theta$ is too small). By contrast, system-dependent schemes—which tailor $\theta$ to the specific electronic structure of the target 
system—generally offer higher accuracy. For example, the self-consistent stability index scheme \cite{lin_self-consistent_2017} dynamically adjusts $\theta$ during the calculation based on a stability index derived from 
the system's orbital occupation characteristics. This approach automatically drives $\theta$ toward zero for SR systems (recovering standard KS-DFT) while increasing $\theta$ for strong polyradicals. Consequently, 
this system-dependent $\theta$-scheme provides a more balanced description of SR and MR systems than system-independent approaches. However, its performance for molecular bond dissociation remains 
unsatisfactory, exhibiting noticeable SCE at the dissociation limits of H$_{2}$ and N$_{2}$. 

Very recently, a response theory \cite{wang_spin_2024} based on TAO-DFT demonstrated that for a molecular system in the lowest singlet state, unphysical spin-symmetry breaking can be completely suppressed by 
employing a sufficiently large fictitious temperature $\theta$. Building on this spin-symmetry restoration criterion, we propose a nonempirical scheme to determine the optimal system-dependent fictitious temperature 
$\theta^{*}$ of a molecular system in its lowest singlet state. 

The remainder of this paper is organized as follows. In \Cref{sec:tao}, we briefly review the formalism of TAO-DFT. In \Cref{sec:theta_star}, we develop our nonempirical scheme for determining the optimal fictitious 
temperatures of molecular systems in their lowest singlet states. In \Cref{sec:results}, we evaluate the performance of the proposed $\theta$-scheme across several MR and SR systems. Finally, we present our 
conclusions in \Cref{sec:conclusions}.

\section{Overview of TAO-DFT}\label{sec:tao} 

\subsection{Self-Consistent Equations} 

Consider the GS of a physical system consisting of $N$ interacting electrons in an external potential $v_{\text{ext}}({\bf r})$ at zero electronic temperature ($\theta_{el} = 0$). In TAO-DFT \cite{chai_density_2012}, 
the GS electron density $\rho({\bf r})$ of the physical system is represented by the thermal equilibrium density $\rho_{s}({\bf r})$ of a reference system (commonly termed the TAO reference system) comprising $N$ 
non-interacting electrons in an effective potential $v^{\theta}_{s}({\bf r})$ at a fictitious temperature $\theta$. For clarity, all $\theta$-dependent quantities (e.g., $v^{\theta}_{s}({\bf r})$) are denoted hereafter by the 
superscript or subscript $\theta$. 

To determine the GS density $\rho({\bf r})$, the self-consistent equations (atomic units are used throughout this work) are given by \cite{chai_density_2012} 
\begin{equation}\label{eq:tao-eq1} 
\bigg\lbrace - \frac{1}{2} {\bf \nabla}^{2} \ + \ v^{\theta}_{s}({\bf r}) \bigg\rbrace \psi^{\theta}_{i}({\bf r}) = \epsilon^{\theta}_{i} \psi^{\theta}_{i}({\bf r}), 
\end{equation} 
where $i$ denotes the spin-orbital index, and the effective potential is defined as 
\begin{equation}\label{eq:tao-eq2} 
v^{\theta}_{s}({\bf r}) = v_{\text{ext}}({\bf r}) + \frac{\delta E_{\text{H}}[\rho]}{\delta \rho({\bf r})} + \frac{\delta E_{\text{xc}}[\rho]}{\delta \rho({\bf r})} + \frac{\delta E_{\theta}[\rho]}{\delta \rho({\bf r})}. 
\end{equation} 
Here, the first term is the external potential, and the second term is the Hartree potential, where 
\begin{equation} 
E_{\text{H}}[\rho] = \frac{1}{2} \iint \frac{\rho({\bf r})\rho({\bf r'})}{|{\bf r} - {\bf r'}|}d{\bf r}d{\bf r'} 
\end{equation} 
is the Hartree energy functional. The third term is the xc potential, where $E_{\text{xc}}[\rho]$ is 
the standard xc energy functional as defined in KS-DFT \cite{kohn_self-consistent_1965}. 
The fourth term is the $\theta$-dependent potential, where the $\theta$-dependent energy functional $E_{\theta}[\rho]$ is defined as 
\begin{equation}\label{eq:Etheta} 
E_{\theta}[\rho] = A_{s}^{\theta=0}[\rho] - A_{s}^{\theta}[\rho], 
\end{equation} 
with $A_{s}^{\theta}[\rho]$ representing the non-interacting kinetic free energy functional (expressed in terms of the density $\rho({\bf r})$) 
at the fictitious temperature $\theta$. By construction, the GS density $\rho({\bf r})$ is represented by $\rho_{s}({\bf r})$: 
\begin{equation}\label{eq:tao-eq3} 
\rho({\bf r}) = \rho_{s}({\bf r}) = \sum_{i=1}^{\infty} f^{\theta}_{i} |\psi^{\theta}_{i}({\bf r})|^{2}. 
\end{equation} 
Here, $f^{\theta}_{i}$ is the occupation number of the $i$th TAO orbital $\psi^{\theta}_{i}({\bf r})$, governed by the FD distribution function 
\begin{equation}\label{eq:tao-eq4} 
f^{\theta}_{i} = \{1 + \text{exp}[(\epsilon^{\theta}_{i} - \mu) / \theta]\}^{-1}, 
\end{equation} 
where $\epsilon^{\theta}_{i}$ is the energy of the $i$th TAO orbital $\psi^{\theta}_{i}({\bf r})$, and $\mu$ is the chemical potential chosen to conserve the total electron number $N$: 
\begin{equation}\label{eq:tao-eq5} 
\sum_{i=1}^{\infty} \{1 + \text{exp}[(\epsilon^{\theta}_{i} - \mu) / \theta]\}^{-1} = N. 
\end{equation} 
The self-consistent procedure defined by \Cref{eq:tao-eq1,eq:tao-eq2,eq:tao-eq3,eq:tao-eq4,eq:tao-eq5} determines 
the TAO orbitals $\{\psi^{\theta}_{i}({\bf r})\}$, the TOONs $\{f^{\theta}_{i}\}$, and the GS density $\rho({\bf r})$ \cite{chai_density_2012}. 

Once self-consistency is achieved, the GS electronic energy $E_{\text{TAO-DFT}}[\rho]$ of the physical system is evaluated as 
\begin{equation}\label{eq:tao-energy} 
E_{\text{TAO-DFT}}[\rho] = U_{\text{ext}}[\rho] + A_{s}^{\theta}[\{f^{\theta}_{i},\psi^{\theta}_{i}\}] + E_{\text{H}}[\rho] + E_{\text{xc}\theta}[\rho], 
\end{equation} 
where 
\begin{equation}\label{eq:ext-energy} 
U_{\text{ext}}[\rho] = \int \rho({\bf r}) v_{\text{ext}}({\bf r})d{\bf r} 
\end{equation} 
is the external potential energy functional, 
\begin{equation}\label{eq:Exctheta} 
E_{\text{xc}\theta}[\rho] = E_{\text{xc}}[\rho] + E_{\theta}[\rho] 
\end{equation} 
is the xc$\theta$ energy functional, and 
\begin{equation}\label{eq:tao-As} 
A_{s}^{\theta}[\{f^{\theta}_{i},\psi^{\theta}_{i}\}] = T_{s}^{\theta}[\{f^{\theta}_{i},\psi^{\theta}_{i}\}] + E_{S}^{\theta}[\{f^{\theta}_{i}\}] 
\end{equation} 
is the non-interacting kinetic free energy (expressed in terms of the TAO orbitals and TOONs) at the fictitious temperature $\theta$. 
This free energy can be computed exactly as the sum of the kinetic energy, 
\begin{equation}\label{eq:tao-Ts} 
T_{s}^{\theta}[\{f^{\theta}_{i},\psi^{\theta}_{i}\}] 
= -\frac{1}{2} \sum_{i=1}^{\infty} f^{\theta}_{i} \int \psi^{\theta*}_{i}({\bf r}){\bf \nabla}^{2}\psi^{\theta}_{i}({\bf r})d{\bf r}, 
\end{equation} 
and the entropy contribution, 
\begin{equation}\label{eq:tao-Es} 
E_{S}^{\theta}[\{f^{\theta}_{i} \}] 
= \theta \sum_{i=1}^{\infty} \bigg\lbrace f^{\theta}_{i}\ \text{ln}(f^{\theta}_{i}) + (1-f^{\theta}_{i})\ \text{ln}(1-f^{\theta}_{i}) \bigg\rbrace, 
\end{equation} 
of non-interacting electrons at the fictitious temperature $\theta$. 

In the limiting case of $\theta = 0$, because $E_{\theta=0}[\rho] = 0$ (see \Cref{eq:Etheta}), 
TAO-DFT \cite{chai_density_2012} with the xc$\theta$ energy functional $E_{\text{xc}\theta}[\rho]$ (see \Cref{eq:Exctheta}) reduces directly to 
KS-DFT \cite{kohn_self-consistent_1965} with the standard xc energy functional $E_{\text{xc}}[\rho]$.

\subsection{Role of the Fictitious Temperature $\theta$} 

Both KS-DFT and TAO-DFT are electronic structure methods used to investigate the GS properties of physical systems at zero electronic temperature ($\theta_{el} = 0$), 
though they differ in several key respects. 

Suppose that the GS density $\rho({\bf r})$ of a physical system is representable in both KS-DFT (i.e., TAO-DFT with $\theta = 0$) and TAO-DFT (with a finite fictitious temperature $\theta$). 
In TAO-DFT \cite{chai_density_2012}, the Hohenberg-Kohn (HK) \cite{hohenberg_inhomogeneous_1964} universal functional $F_{\text{HK}}[\rho]$—defined as the sum of the interacting 
kinetic energy $T[\rho]$ and the electron–electron repulsion energy $V_{ee}[\rho]$ at zero electronic temperature ($\theta_{el} = 0$)—is expressed as 
\begin{equation} 
F_{\text{HK}}[\rho] = A_{s}^{\theta}[\rho] + E_{\text{H}}[\rho] + E_{\text{xc}\theta}[\rho], 
\end{equation} 
where 
$E_{\text{xc}\theta}[\rho] \equiv F_{\text{HK}}[\rho] - A_{s}^{\theta}[\rho] - E_{\text{H}}[\rho]$ is the xc$\theta$ energy functional, and 
$A_{s}^{\theta}[\rho] = A_{s}^{\theta}[\{f^{\theta}_{i},\psi^{\theta}_{i}\}]$ (see \Cref{eq:tao-As}) is the non-interacting kinetic free energy at the fictitious temperature $\theta$, 
evaluated explicitly using the TAO orbitals $\{\psi^{\theta}_{i}({\bf r})\}$ and TOONs $\{f^{\theta}_{i}\}$. 

By contrast, in KS-DFT \cite{kohn_self-consistent_1965}, the HK universal functional $F_{\text{HK}}[\rho]$ is given by 
\begin{equation} 
F_{\text{HK}}[\rho] = T_{s}[\rho] + E_{\text{H}}[\rho] + E_{\text{xc}}[\rho], 
\end{equation} 
where 
$E_{\text{xc}}[\rho] \equiv F_{\text{HK}}[\rho] - T_{s}[\rho] - E_{\text{H}}[\rho]$ is the standard xc energy functional, and 
$T_{s}[\rho] = T_{s}[\{\phi_{i}\}]$ is the non-interacting kinetic energy at $\theta = 0$ (i.e., $T_{s}[\rho] = A_{s}^{\theta=0}[\rho]$), 
evaluated explicitly with the occupied Kohn-Sham (KS) orbitals $\{\phi_{i}({\bf r})\}$: 
\begin{equation} 
T_{s}[\{\phi_{i}\}] = -\frac{1}{2} \sum_{i=1}^{N} \int \phi^{*}_{i}({\bf r}){\bf \nabla}^{2}\phi_{i}({\bf r})d{\bf r}. 
\end{equation} 
Furthermore, the GS density $\rho({\bf r})$ is represented by the occupied KS orbitals $\{\phi_{i}({\bf r})\}$ as 
\begin{equation}\label{eq:KS-desnity} 
\rho({\bf r}) = \sum_{i=1}^{N} |\phi_{i}({\bf r})|^{2}. 
\end{equation} 

Therefore, when the GS density $\rho({\bf r})$ of a physical system is representable in both KS-DFT ($\theta = 0$) and TAO-DFT (with fictitious temperature $\theta$), 
the exact GS electronic energy $E_{0}$—given by the sum of the external potential energy functional $U_{\text{ext}}[\rho]$ (see \Cref{eq:ext-energy}) and the HK universal functional 
$F_{\text{HK}}[\rho]$ evaluated at $\rho({\bf r})$—is identical in both frameworks by construction: 
\begin{equation} 
E_{0} = E_{\text{TAO-DFT}}[\rho] = E_{\text{KS-DFT}}[\rho], 
\end{equation} 
where $E_{\text{TAO-DFT}}[\rho]$ (see \Cref{eq:tao-energy}) and $E_{\text{KS-DFT}}[\rho]$ denote the GS electronic energies in TAO-DFT and KS-DFT, respectively. 

However, the GS density $\rho({\bf r})$ of an MR system may not be representable within KS-DFT ($\theta = 0$) \cite{baerends,KS-rep-1,KS-rep-2,KS-rep-3}. In such cases, the exact 
GS density $\rho({\bf r})$ and electronic energy $E_{0}$ are inaccessible within KS-DFT. TAO-DFT overcomes this limitation by introducing a tunable fictitious temperature $\theta$ to 
enhance GS density representability. Specifically, the GS density $\rho({\bf r})$ of any physical system (whether SR or MR) at $\theta_{el} = 0$ can be expressed in terms of its exact 
natural orbitals (NOs) $\{\chi_{i}({\bf r})\}$ and NOONs $\{n_{i}\}$ \cite{NO}: 
\begin{equation}\label{eq:FT-RDM-desnity} 
\rho({\bf r}) = \sum_{i=1}^{\infty} n_{i} |\chi_{i}({\bf r})|^{2}. 
\end{equation} 
By selecting an appropriate $\theta$ in TAO-DFT, the resulting TOONs $\{f^{\theta}_{i}\}$ and TAO orbitals $\{\psi^{\theta}_{i}({\bf r})\}$ closely approximate the exact NOONs $\{n_{i}\}$ and 
NOs $\{\chi_{i}({\bf r})\}$, respectively. Owing to the analogous mathematical forms of their density representations (see \Cref{eq:tao-eq3,eq:FT-RDM-desnity}), a suitably tailored $\theta$ can 
significantly improve GS density representability. As discussed in Section III.E of Chai \cite{chai_density_2012}, with this tailored $\theta$, the interacting kinetic energy (expressed in terms 
of the NOs and NOONs), 
\begin{equation}\label{eq:exact-T} 
T[\rho] = T[\{n_{i},\chi_{i}\}] = -\frac{1}{2} \sum_{i=1}^{\infty} n_{i} \int \chi^{*}_{i}({\bf r}){\bf \nabla}^{2}\chi_{i}({\bf r})d{\bf r}, 
\end{equation} 
can be approximated by the non-interacting kinetic energy $T_{s}^{\theta}[\{f^{\theta}_{i},\psi^{\theta}_{i}\}]$: 
\begin{equation} 
T[\{n_{i},\chi_{i}\}] \approx T_{s}^{\theta}[\{f^{\theta}_{i},\psi^{\theta}_{i}\}], 
\end{equation} 
owing to the analogous mathematical forms of their energy expressions (see \Cref{eq:tao-Ts,eq:exact-T}). Accordingly, the electron–electron repulsion energy, 
\begin{equation} 
\begin{aligned} 
V_{ee}[\rho] 
&= F_{\text{HK}}[\rho] - T[\rho]  \\ 
&= (A_{s}^{\theta}[\{f^{\theta}_{i},\psi^{\theta}_{i}\}] + E_{\text{H}}[\rho] + E_{\text{xc}\theta}[\rho]) - T[\{n_{i},\chi_{i}\}]  \\ 
&= (T_{s}^{\theta}[\{f^{\theta}_{i},\psi^{\theta}_{i}\}] + E_{S}^{\theta}[\{f^{\theta}_{i}\}] + E_{\text{H}}[\rho] + E_{\text{xc}\theta}[\rho]) - T[\{n_{i},\chi_{i}\}]  \\ 
&= E_{\text{H}}[\rho] + E_{\text{xc}\theta}[\rho] + E_{S}^{\theta}[\{f^{\theta}_{i}\}] + (T_{s}^{\theta}[\{f^{\theta}_{i},\psi^{\theta}_{i}\}] - T[\{n_{i},\chi_{i}\}]), 
\end{aligned} 
\end{equation} 
can be approximated as 
\begin{equation}\label{eq:Vee} 
V_{ee}[\rho] \approx E_{\text{H}}[\rho] + E_{\text{xc}\theta}[\rho] + E_{S}^{\theta}[\{f^{\theta}_{i}\}], 
\end{equation} 
which naturally incorporates the entropy contribution $E_{S}^{\theta}[\{f^{\theta}_{i}\}]$ (see \Cref{eq:tao-Es}). 
In particular, as shown in TAO-DFT \cite{chai_density_2012}, this entropy contribution provides a reliable approximation to the static correlation energy. 
Consequently, the optimal $\theta$ should remain vanishingly small for SR systems, whereas it must be tailored to each specific MR system.

\subsection{Role of the $\theta$-Dependent Energy Functional $E_{\theta}[\rho]$} 

Here, we examine the theoretical necessity of the $\theta$-dependent energy functional $E_{\theta}[\rho]$ in TAO-DFT \cite{chai_density_2012}. Within the exact TAO-DFT 
formalism, the GS electronic energy $E_{\text{TAO-DFT}}[\rho]$ (see \Cref{eq:tao-energy}) is strictly independent of the fictitious temperature $\theta$, because the sum 
\begin{equation} 
A_{s}^{\theta}[\{f^{\theta}_{i},\psi^{\theta}_{i}\}] + E_{\theta}[\rho] 
= A_{s}^{\theta}[\rho] + (A_{s}^{\theta=0}[\rho] - A_{s}^{\theta}[\rho]) 
= A_{s}^{\theta=0}[\rho] 
\end{equation} 
exactly cancels all $\theta$-dependence. Consequently, neglecting $E_{\theta}[\rho]$ introduces a spurious $\theta$-dependence into $E_{\text{TAO-DFT}}[\rho]$, except in 
the trivial limit $\theta = 0$, where TAO-DFT reduces directly to standard KS-DFT. 

Furthermore, omitting $E_{\theta}[\rho]$ reduces the effective potential $v^{\theta}_{s}({\bf r})$ (see \Cref{eq:tao-eq2}) to the $\theta$-independent KS potential 
$v^{\text{KS}}_{s}({\bf r})$ \cite{kohn_self-consistent_1965}: 
\begin{equation}\label{eq:ks-potential2} 
v^{\theta}_{s}({\bf r}) \approx v^{\text{KS}}_{s}({\bf r}) = v_{\text{ext}}({\bf r}) + \frac{\delta E_{\text{H}}[\rho]}{\delta \rho({\bf r})} + \frac{\delta E_{\text{xc}}[\rho]}{\delta \rho({\bf r})}. 
\end{equation} 
Substituting the exact GS density $\rho({\bf r})$ into $v^{\text{KS}}_{s}({\bf r})$ yields $\theta$-independent KS orbitals $\{\phi_{i}({\bf r})\}$ and orbital energies $\{\epsilon_{i}\}$. 
However, the chemical potential $\mu^{\theta}$, which is determined by conserving the total electron number $N$, retains an explicit $\theta$-dependence: 
\begin{equation}
\sum_{i=1}^{\infty} \{1 + \text{exp}[(\epsilon_{i} - \mu^{\theta}) / \theta]\}^{-1} = N. 
\end{equation} 
Because the resulting orbital occupation numbers $\{{\tilde f}^{\theta}_{i}\}$ remain explicitly dependent on $\theta$, 
\begin{equation} 
{\tilde f}^{\theta}_{i} = \{1 + \text{exp}[(\epsilon_{i} - \mu^{\theta}) / \theta]\}^{-1}, 
\end{equation} 
the reconstructed GS density also becomes spuriously $\theta$-dependent: 
\begin{equation} 
\rho^{\theta}({\bf r}) = \sum_{i=1}^{\infty} {\tilde f}^{\theta}_{i} |\phi_{i}({\bf r})|^{2}. 
\end{equation} 
For any $\theta > 0$, this $\theta$-dependent density $\rho^{\theta}({\bf r})$ contradicts the exact, $\theta$-independent GS density $\rho({\bf r})$ initially substituted into the potential. 

Therefore, $E_{\theta}[\rho]$ cannot be formally neglected; instead, it serves as an essential mathematical bridge in TAO-DFT, enabling the use of fractional orbital occupations (via $\theta$) to 
capture strong static correlation while guaranteeing that—within the exact formalism—both the resulting GS density $\rho({\bf r})$ (see \Cref{eq:tao-eq3}) and GS electronic energy 
$E_{\text{TAO-DFT}}[\rho]$ (see \Cref{eq:tao-energy}) remain exact and invariant to $\theta$.

\section{Nonempirical Determination of the Optimal Fictitious Temperature in TAO-DFT via Spin-Symmetry Restoration}\label{sec:theta_star} 

\subsection{Spin-Symmetry Restoration Scheme} 

As noted above, the fictitious temperature $\theta$ in TAO-DFT \cite{chai_density_2012} should, in principle, be chosen such that the TOONs closely approximate the exact NOONs of the system \cite{NO}. 
This alignment significantly improves GS density representability and enables the entropy contribution to accurately capture the static correlation energy (see, e.g., Section III.E of Chai \cite{chai_density_2012}). 
In practice, however, this approach is rarely feasible because computing exact NOONs beforehand defeats the low-cost advantage of TAO-DFT. 

To circumvent this limitation, we propose a nonempirical scheme to determine the optimal fictitious temperature $\theta^{*}$ for a molecular system in its lowest singlet state, demonstrating that the resulting 
TOONs qualitatively match the NOONs obtained from accurate MR electronic structure methods. 

For a molecular system in the lowest singlet state, the spin-symmetry constraint requires that the spin-restricted and spin-unrestricted energies obtained from an exact theory must be 
identical \cite{cohen_insights_2008,Yang12,DFT-exchange,chai_density_2012,wang_spin_2024}. Accordingly, for a given electronic structure method, a quantitative measure of the SCE can be defined via 
the restricted–unrestricted (RU) energy difference: 
\begin{equation}\label{eq:RUG} 
\Delta E_{\text{RU}} = E_{\text{RS}} - E_{\text{US}}, 
\end{equation} 
where $E_{\text{RS}}$ and $E_{\text{US}}$ denote the spin-restricted and spin-unrestricted energies, respectively, of the lowest singlet state of the molecular system. By construction, the RU energy difference 
is non-negative ($\Delta E_{\text{RU}} \ge 0$). For an exact theory, the spin-symmetry constraint 
\begin{equation}\label{eq:SSC} 
\Delta E_{\text{RU}} = 0 
\end{equation} 
must be satisfied for both SR and MR systems. In KS-DFT with conventional xc functionals, $\Delta E_{\text{RU}}$ is expected to be vanishingly small for SR systems (e.g., closed-shell molecules with 
sizeable HOMO–LUMO gaps), whereas it can become considerably large for MR systems (e.g., during molecular bond dissociation) due to severe spin-symmetry breaking \cite{Yang12,DFT-exchange}. 
Recently, Wang and Chai \cite{wang_spin_2024} demonstrated that with a sufficiently large $\theta$, TAO-DFT employing an xc$\theta$ functional eliminates unphysical spin-symmetry breaking in the 
lowest singlet state of MR systems, yielding vanishingly small SCEs. 

Following Wang and Chai \cite{wang_spin_2024}, we define the optimal fictitious temperature ($\theta^{*}$) for the lowest singlet state of a molecular system in TAO-DFT with an xc$\theta$ functional as 
the minimum $\theta$ at which $\Delta E_{\text{RU}}$ becomes vanishingly small—that is, achieving spin-symmetry restoration within a numerical tolerance $\varepsilon_{\text{RU}}$ 
($\Delta E_{\text{RU}} < \varepsilon_{\text{RU}}$). Based on this spin-symmetry restoration criterion, the optimal $\theta^{*}$ values for both SR and MR systems can be determined nonempirically. Notably, 
TAO-DFT with an xc$\theta$ functional reduces at $\theta = 0$ to KS-DFT with the corresponding xc functional, which already yields a vanishingly small $\Delta E_{\text{RU}}$ for SR systems. Consequently, 
the present scheme is anticipated to yield a vanishingly small $\theta^{*}$ for SR systems. By employing $\theta^{*}$ in TAO-DFT, unphysical spin-symmetry breaking in MR systems is effectively suppressed, 
while spurious entropy contributions in SR systems are largely avoided.

\subsection{Computational Details} 

We evaluate the performance of TAO-LDA (i.e., TAO-DFT with the LDA xc$\theta$ functional) \cite{chai_density_2012} within the proposed $\theta$-scheme for the lowest singlet states of several molecular 
systems, encompassing both MR systems (the bond dissociation of H$_{2}$, N$_{2}$, Li$_{2}$, and LiH, as well as the torsional rotation of twisted ethylene) and SR systems (e.g., the dissociation of He$_{2}$ 
and Ne$_{2}$). These results are compared against those obtained using TAO-LDA with various fixed $\theta$ values ($\theta$ = 0, 7, 10, 15, 20, 30, and 40 mHartree). 

Numerically, KS-LDA (i.e., KS-DFT with the LDA xc energy functional) \cite{LDA-1,LDA-2}, which corresponds to TAO-LDA at $\theta = 0$, can yield multiple self-consistent field (SCF) solutions along 
molecular dissociation pathways \cite{multiple1}. This behavior stems primarily from the non-linear nature of the SCF equations; as bonds are stretched or broken, the electronic structure transitions from SR 
to MR character, giving rise to multiple electronic solutions. While TAO-LDA with $\theta > 0$ mitigates this issue, multiple electronic solutions cannot be completely eliminated unless $\theta$ is sufficiently 
large \cite{chai_density_2012,wang_spin_2024}. 

To efficiently locate $\theta^{*}$ despite the presence of multiple SCF solutions at $\theta = 0$ and small $\theta$ values, we determine $\theta^{*}$ via a two-stage $\theta$-scan based on spin-symmetry 
restoration within a numerical tolerance of $\varepsilon_{\text{RU}} = 0.1$ mHartree. Adopting a tighter $\varepsilon_{\text{RU}}$ enforces the spin-symmetry constraint of $\Delta E_{\text{RU}} = 0$ (see 
\Cref{eq:SSC}) more strictly, albeit at the increased computational cost of converging the true spin-restricted and spin-unrestricted energies to tighter numerical precision. In the first stage, a coarse 
$\theta$-scan is conducted from 0 to 40 mHartree with a step size of 0.1 mHartree to identify $\theta^{\dagger}$ (the minimum $\theta$ satisfying $\Delta E_{\text{RU}} < \varepsilon_{\text{RU}}$ in the first 
stage). If $\theta^{\dagger} = 0$, then $\theta^{*} = 0$. If $\theta^{\dagger} > 0$, a finer second-stage $\theta$-scan is performed from ($\theta^{\dagger} - 0.1$ mHartree) to $\theta^{\dagger}$ with a step size 
of 0.01 mHartree to obtain $\theta^{*}$ (the minimum $\theta$ satisfying the criterion in the second stage). 

All calculations are performed using \textsf{Q-Chem 6.1.1} \cite{epifanovsky_software_2021}. For the MR systems (bond-breaking and bond-twisting systems), all TAO-LDA calculations employ the 6-31G(d) 
basis set and the EML(75,302) integration grid, consisting of 75 Euler–Maclaurin radial grid points and 302 Lebedev angular grid points. For the SR systems (weakly bound complexes), all TAO-LDA 
calculations employ the aug-cc-pVQZ basis set and the EML(250,590) grid. For all systems studied, reference NOONs are computed using the ASCI-SCF method—namely, the complete active space 
self-consistent field (CASSCF) method utilizing the adaptive sampling configuration interaction (ASCI) algorithm as its active-space solver \cite{asci}—with the aug-cc-pVQZ basis set.

\section{Results and discussion}\label{sec:results} 

\subsection{MR systems} 

\subsubsection{H$_{2}$ dissociation} 

The dissociation of H$_{2}$ (a single-bond-breaking system) presents a well-known challenge for KS-DFT, as strong static correlation effects become significant in the bond-stretched region. Due to the spin-symmetry 
constraint, the spin-restricted and spin-unrestricted potential energy curves of H$_{2}$ obtained from an exact theory must be identical. In other words, $\Delta E_{\text{RU}} = 0$ (see \Cref{eq:SSC}) must be satisfied 
at all internuclear distances. Beyond the Coulson-Fischer (CF) point \cite{zhang_unmasking_2020}, however, the spin-restricted and spin-unrestricted energies of KS-DFT diverge, yielding 
$\Delta E_{\text{RU}} > 0$ \cite{DFT-exchange,cohen_insights_2008,Yang12}. Therefore, we adopt the difference between the spin-restricted and spin-unrestricted dissociation limits ($\Delta E_{\text{RU}}$ at the 
dissociation limit) obtained from an approximate electronic structure method as a quantitative measure of its SCE. Owing to spin-symmetry restoration, TAO-LDA employing the optimal $\theta^{*}$ values guarantees 
a vanishingly small $\Delta E_{\text{RU}}$ ($\Delta E_{\text{RU}} < \varepsilon_{\text{RU}}$) at all internuclear distances and thus essentially eliminates the SCE. 

\begin{figure} 
\includegraphics[scale=1.12]{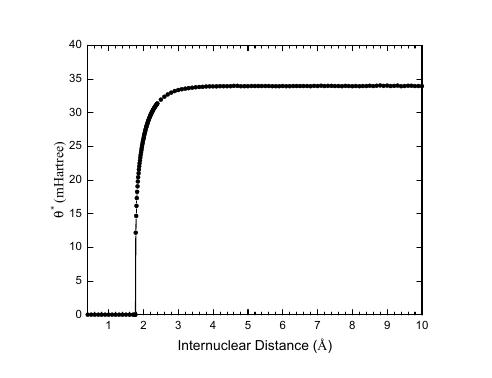} 
\caption{\label{fgr:h2_theta_star} 
Optimal $\theta^{*}$ values for the lowest singlet state of H$_{2}$ as a function of the internuclear distance $R$, calculated using TAO-LDA within the present scheme.} 
\end{figure}

Applying the present scheme, we determine the optimal $\theta^{*}$ values (see \Cref{fgr:h2_theta_star}) for the lowest singlet state of H$_{2}$ as a function of the internuclear distance $R$, calculated using TAO-LDA. 
Near the equilibrium bond length, $\theta^{*}$ remains zero; here, TAO-LDA reduces to standard KS-LDA ($\theta = 0$) and accurately describes the SR character. Once $R$ exceeds approximately 1.78 {\AA} (the CF 
point of KS-LDA), $\theta^{*}$ increases rapidly—signaling a transition from SR to MR character—before approaching a plateau of approximately 34 mHartree at the dissociation limit, where MR character is pronounced. 

\begin{figure} 
\includegraphics[scale=1.12]{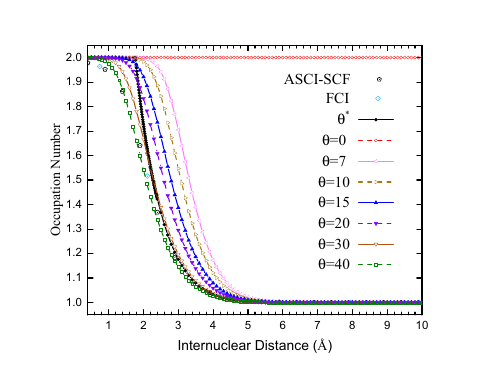} 
\caption{\label{fgr:h2_homo} 
Occupation numbers of the 1$\sigma_{g}$ orbital for the lowest singlet state of H$_{2}$ as a function of the internuclear distance $R$, calculated using spin-restricted TAO-LDA with the optimal 
$\theta^{*}$ values (see \Cref{fgr:h2_theta_star}) and various fixed $\theta$ values (in mHartree). The $\theta = 0$ case corresponds to spin-restricted KS-LDA. 
The reference data are the NOONs of FCI (taken from Ref.\ \cite{h2mrci}) and ASCI-SCF (obtained in this work using the method of Ref.\ \cite{asci}).} 
\end{figure}

To assess whether the present scheme properly captures the evolving MR character during bond dissociation, we examine the occupation numbers of the 1$\sigma_{g}$ orbital (HOMO) for the lowest singlet state of 
H$_{2}$ across varying $R$, calculated via spin-restricted TAO-LDA with various $\theta$ values (see \Cref{fgr:h2_homo}). The reference data comprise the NOONs from full configuration interaction (FCI) \cite{h2mrci} 
and ASCI-SCF \cite{asci} calculations, with the ASCI-SCF calculations employing an active space of 2 electrons in 92 orbitals (expanding the wavefunction across 8,464 determinants). Near the equilibrium bond length, 
the 1$\sigma_{g}$ occupation numbers of spin-restricted TAO-LDA (with $\theta^{*}$) equal 2.0 (doubly occupied). They drop rapidly around $R = 1.78$ {\AA} before approaching 1.0 (singly occupied) at the dissociation 
limit. Overall, the 1$\sigma_{g}$ occupation numbers of spin-restricted TAO-LDA (with $\theta^{*}$) match reasonably well with the FCI and ASCI-SCF NOONs. 

\begin{figure} 
\includegraphics[scale=1.12]{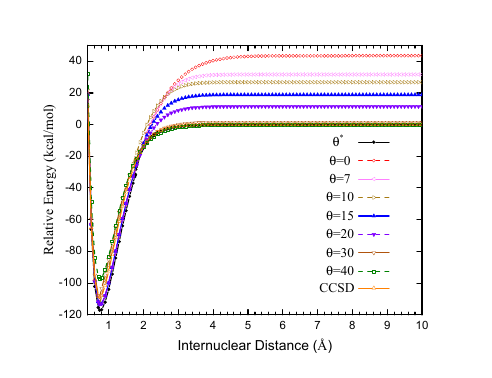} 
\caption{\label{fgr:h2_relative} 
Potential energy curves (in relative energy) for the lowest singlet state of H$_{2}$ as a function of the internuclear distance $R$, calculated using spin-restricted TAO-LDA with the optimal $\theta^{*}$ 
values (see \Cref{fgr:h2_theta_star}) and various fixed $\theta$ values (in mHartree). The $\theta = 0$ case corresponds to spin-restricted KS-LDA. The reference potential energy curve is calculated 
using CCSD theory \cite{ccsd}. The zeros of energy are set at the respective spin-unrestricted dissociation limits.} 
\end{figure}

To evaluate how the present scheme addresses SCE, \Cref{fgr:h2_relative} presents the potential energy curves (in relative energy) for the lowest singlet state of H$_{2}$ as a function of $R$, calculated using 
spin-restricted TAO-LDA with various $\theta$ values, where the zeros of energy are set at the respective spin-unrestricted dissociation limits. The reference curve is computed using coupled-cluster theory with iterative 
singles and doubles (CCSD)—an exact method for any two-electron system \cite{ccsd}—employing the aug-cc-pVQZ basis set. TAO-LDA (with $\theta^{*}$) performs very similarly to KS-LDA ($\theta = 0$) near the 
equilibrium bond length (the SR region), and yields a vanishingly small SCE at the dissociation limit (the MR region), agreeing reasonably well with the exact CCSD curve. 

\begin{figure} 
\includegraphics[scale=1.12]{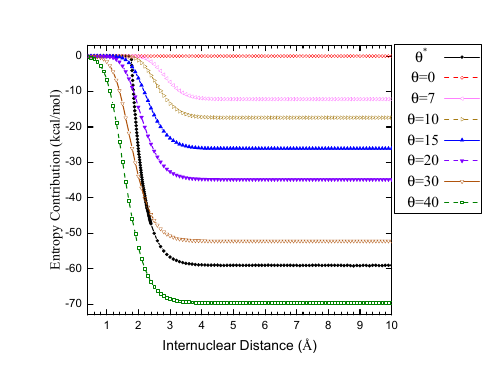} 
\caption{\label{fgr:h2_entropy} 
Entropy contributions (approximations to the static correlation energy) for the lowest singlet state of H$_{2}$ as a function of the internuclear distance $R$, calculated using spin-restricted TAO-LDA 
with the optimal $\theta^{*}$ values (see \Cref{fgr:h2_theta_star}) and various fixed $\theta$ values (in mHartree). The $\theta = 0$ case corresponds to spin-restricted KS-LDA.} 
\end{figure}

To explore how the entropy contribution $E_{S}^{\theta}[\{f^{\theta}_{i}\}]$ (see \Cref{eq:tao-Es}) evolves during bond dissociation, we plot the entropy contributions (approximations to the static correlation energy) for the 
lowest singlet state of H$_{2}$ as a function of $R$, computed using spin-restricted TAO-LDA with various $\theta$ values (see \Cref{fgr:h2_entropy}). The entropy contribution of spin-restricted TAO-LDA (with 
$\theta^{*}$) is negligible near the equilibrium bond length, becomes pronounced once $R$ exceeds approximately 1.78 {\AA}, and approaches a negative constant at the dissociation limit, properly simulating the static 
correlation energy throughout bond dissociation.

\subsubsection{N$_{2}$ dissociation} 

\begin{figure} 
\includegraphics[scale=1.12]{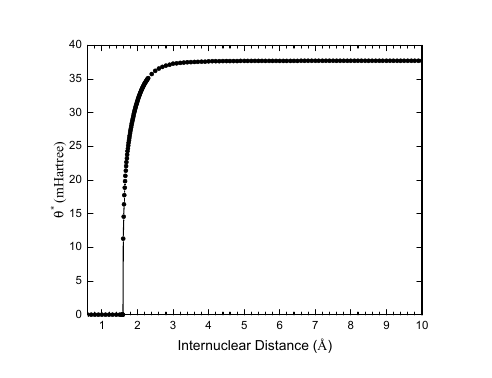} 
\caption{\label{fgr:n2_theta} 
Optimal $\theta^{*}$ values for the lowest singlet state of N$_{2}$ as a function of the internuclear distance $R$, calculated using TAO-LDA within the present scheme.} 
\end{figure}

Similar trends are observed for N$_{2}$ dissociation, a representative triple-bond-breaking system. \Cref{fgr:n2_theta} plots the optimal $\theta^{*}$ values for the lowest singlet state of N$_{2}$ as a function of the 
internuclear distance $R$, computed via TAO-LDA. Near the equilibrium bond length (the SR region), $\theta^{*}$ remains zero. Once $R$ exceeds approximately 1.6 {\AA}—the CF point of KS-LDA—$\theta^{*}$ rises 
sharply, signaling a transition from SR to MR character. At the dissociation limit (the MR region), $\theta^{*}$ plateaus at approximately 38 mHartree. 

\begin{figure} 
\includegraphics[scale=1.12]{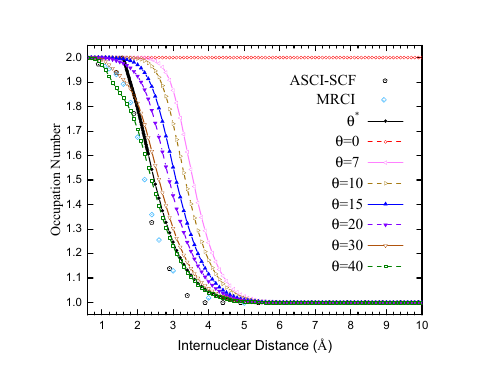} 
\caption{\label{fgr:n2_3sg} 
Occupation numbers of the 3$\sigma_{g}$ orbital for the lowest singlet state of N$_{2}$ as a function of the internuclear distance $R$, calculated using spin-restricted TAO-LDA with the optimal 
$\theta^{*}$ values (see \Cref{fgr:n2_theta}) and various fixed $\theta$ values (in mHartree). The $\theta = 0$ case corresponds to spin-restricted KS-LDA. 
The reference data are the NOONs of MRCI (taken from Ref.\ \cite{n2mrci}) and ASCI-SCF (obtained in this work using the method of Ref.\ \cite{asci}).} 
\end{figure}

\begin{figure} 
\includegraphics[scale=1.12]{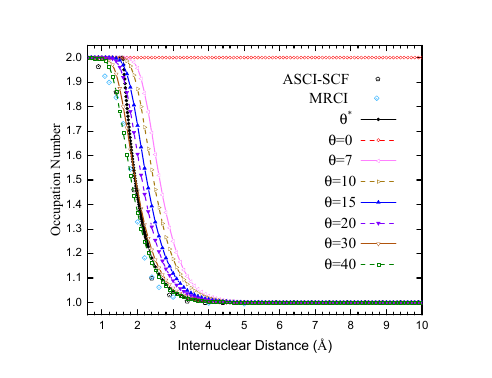} 
\caption{\label{fgr:n2_1pux} 
Occupation numbers of the 1$\pi_{ux}$ orbital for the lowest singlet state of N$_{2}$ as a function of the internuclear distance $R$, calculated using spin-restricted TAO-LDA with the optimal 
$\theta^{*}$ values (see \Cref{fgr:n2_theta}) and various fixed $\theta$ values (in mHartree). The $\theta = 0$ case corresponds to spin-restricted KS-LDA. 
The reference data are the NOONs of MRCI (taken from Ref.\ \cite{n2mrci}) and ASCI-SCF (obtained in this work using the method of Ref.\ \cite{asci}).} 
\end{figure}

To verify that our scheme captures the evolving MR character during bond stretching, we examine the occupation numbers of the 3$\sigma_{g}$ and 1$\pi_{ux}$ orbitals (\Cref{fgr:n2_3sg} and \Cref{fgr:n2_1pux}, 
respectively) for the lowest singlet state of N$_{2}$ across varying $R$, calculated via spin-restricted TAO-LDA with various $\theta$ values. The reference data comprise the NOONs from both MR configuration 
interaction (MRCI) \cite{n2mrci} and ASCI-SCF \cite{asci} calculations. Specifically, the ASCI-SCF reference calculations employ spin purification with an active space of 10 electrons in 15 orbitals, expanding the 
wavefunction across 2 million determinants. Near the equilibrium bond length, the 3$\sigma_{g}$ and 1$\pi_{ux}$ occupation numbers of spin-restricted TAO-LDA (with $\theta^{*}$) remain at 2.0. Both drop 
precipitously around $R = 1.6$ {\AA} before approaching 1.0 at the dissociation limit. Overall, the 3$\sigma_{g}$ and 1$\pi_{ux}$ occupation numbers of spin-restricted TAO-LDA (with $\theta^{*}$) are in reasonably 
good agreement with the MRCI and ASCI-SCF NOONs. 

\begin{figure} 
\includegraphics[scale=1.12]{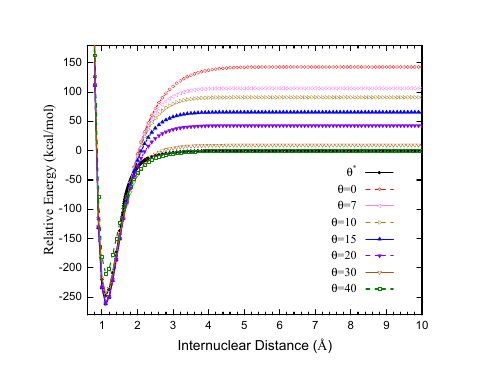} 
\caption{\label{fgr:n2_relative} 
Potential energy curves (in relative energy) for the lowest singlet state of N$_{2}$ as a function of the internuclear distance $R$, calculated using spin-restricted TAO-LDA with the optimal $\theta^{*}$ 
values (see \Cref{fgr:n2_theta}) and various fixed $\theta$ values (in mHartree). The $\theta = 0$ case corresponds to spin-restricted KS-LDA. 
The zeros of energy are set at the respective spin-unrestricted dissociation limits.} 
\end{figure}

\Cref{fgr:n2_relative} displays the potential energy curves (in relative energy) for the lowest singlet state of N$_{2}$ as a function of $R$, computed using spin-restricted TAO-LDA with various $\theta$ values, where 
the zeros of energy are set at the respective spin-unrestricted dissociation limits. TAO-LDA (with $\theta^{*}$) performs very similarly to KS-LDA ($\theta = 0$) near the equilibrium bond length, and yields a vanishingly 
small SCE at the dissociation limit. 

\begin{figure} 
\includegraphics[scale=1.12]{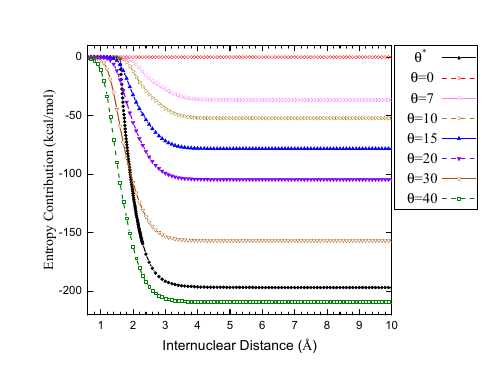} 
\caption{\label{fgr:n2_entropy} 
Entropy contributions (approximations to the static correlation energy) for the lowest singlet state of N$_{2}$ as a function of the internuclear distance $R$, calculated using spin-restricted TAO-LDA 
with the optimal $\theta^{*}$ values (see \Cref{fgr:n2_theta}) and various fixed $\theta$ values (in mHartree). The $\theta = 0$ case corresponds to spin-restricted KS-LDA.} 
\end{figure}

\Cref{fgr:n2_entropy} illustrates the entropy contributions (approximations to the static correlation energy) for the lowest singlet state of N$_{2}$ as a function of $R$, computed using spin-restricted TAO-LDA with 
various $\theta$ values. Analogous to H$_{2}$ dissociation, the entropy contribution of spin-restricted TAO-LDA (with $\theta^{*}$) is negligible near the equilibrium bond length, becomes pronounced once $R$ 
exceeds approximately 1.6 {\AA}, and approaches a negative constant at the dissociation limit (the MR region).

\subsubsection{Li$_{2}$ dissociation} 

\begin{figure} 
\includegraphics[scale=1.12]{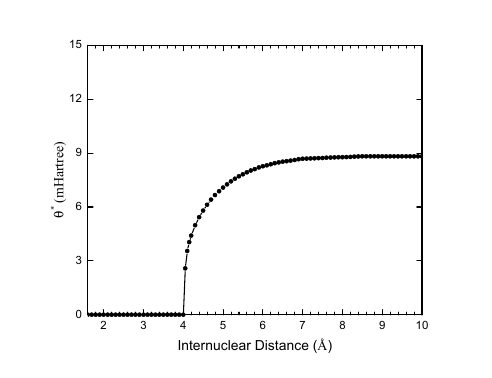} 
\caption{\label{fgr:li2_theta} 
Optimal $\theta^{*}$ values for the lowest singlet state of Li$_{2}$ as a function of the internuclear distance $R$, calculated using TAO-LDA within the present scheme.} 
\end{figure}

Following the same procedure as for the preceding systems, we next examine Li$_{2}$ dissociation, another single-bond-breaking system. We first determine the optimal $\theta^{*}$ values for the lowest singlet state 
of Li$_{2}$ as a function of the internuclear distance $R$, computed using TAO-LDA (see \Cref{fgr:li2_theta}). Near the equilibrium bond length (the SR region), $\theta^{*}$ remains zero. When $R$ exceeds 
approximately 4 {\AA} (the CF point of KS-LDA), $\theta^{*}$ increases sharply, indicating a transition from SR to MR character. At the dissociation limit (the MR region), $\theta^{*}$ plateaus at approximately 9 mHartree. 

\begin{figure} 
\includegraphics[scale=1.12]{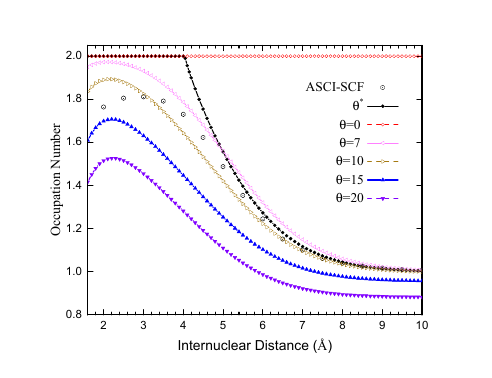} 
\caption{\label{fgr:li2_homo} 
Occupation numbers of the 2$\sigma_{g}$ orbital for the lowest singlet state of Li$_{2}$ as a function of the internuclear distance $R$, calculated using spin-restricted TAO-LDA with the optimal 
$\theta^{*}$ values (see \Cref{fgr:li2_theta}) and various fixed $\theta$ values (in mHartree). The $\theta = 0$ case corresponds to spin-restricted KS-LDA. 
The reference data are the NOONs of ASCI-SCF (obtained in this work using the method of Ref.\ \cite{asci}).} 
\end{figure}

\Cref{fgr:li2_homo} plots the 2$\sigma_{g}$ (HOMO) occupation numbers obtained from spin-restricted TAO-LDA with various $\theta$ values, compared against reference NOONs from ASCI-SCF \cite{asci} calculations. 
Here, the ASCI-SCF calculations employ an active space of 6 electrons in 20 orbitals, expanding the wavefunction across 1,299,600 determinants. Near the equilibrium bond length, the 2$\sigma_{g}$ occupation 
numbers of spin-restricted TAO-LDA (with $\theta^{*}$) remain at 2.0. They drop rapidly around $R = 4$ {\AA} before approaching 1.0 at the dissociation limit. Overall, the 2$\sigma_{g}$ occupation numbers of 
spin-restricted TAO-LDA (with $\theta^{*}$) agree reasonably well with the ASCI-SCF NOONs. 

\begin{figure} 
\includegraphics[scale=1.12]{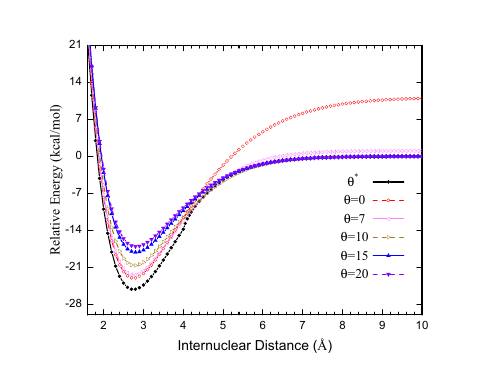} 
\caption{\label{fgr:li2_relative} 
Potential energy curves (in relative energy) for the lowest singlet state of Li$_{2}$ as a function of the internuclear distance $R$, calculated using spin-restricted TAO-LDA with the optimal $\theta^{*}$ 
values (see \Cref{fgr:li2_theta}) and various fixed $\theta$ values (in mHartree). The $\theta = 0$ case corresponds to spin-restricted KS-LDA. 
The zeros of energy are set at the respective spin-unrestricted dissociation limits.} 
\end{figure}

We further examine the potential energy curves (in relative energy) for the lowest singlet state of Li$_{2}$ as a function of $R$, computed using spin-restricted TAO-LDA with various $\theta$ values, where the zeros 
of energy are set at the respective spin-unrestricted dissociation limits. As shown in \Cref{fgr:li2_relative}, TAO-LDA (with $\theta^{*}$) performs very similarly to KS-LDA ($\theta = 0$) near the equilibrium bond length, 
and yields a vanishingly small SCE at the dissociation limit. 

\begin{figure} 
\includegraphics[scale=1.12]{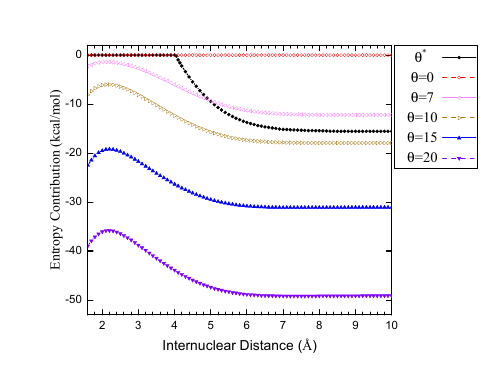} 
\caption{\label{fgr:li2_entropy} 
Entropy contributions (approximations to the static correlation energy) for the lowest singlet state of Li$_{2}$ as a function of the internuclear distance $R$, calculated using spin-restricted TAO-LDA 
with the optimal $\theta^{*}$ values (see \Cref{fgr:li2_theta}) and various fixed $\theta$ values (in mHartree). The $\theta = 0$ case corresponds to spin-restricted KS-LDA.} 
\end{figure}

Finally, \Cref{fgr:li2_entropy} reveals the entropy contributions (approximations to the static correlation energy) for the lowest singlet state of Li$_{2}$ as a function of $R$, computed using spin-restricted TAO-LDA 
with various $\theta$ values. Consistent with the previously examined systems, the entropy contribution of spin-restricted TAO-LDA (with $\theta^{*}$) is negligible near the equilibrium bond length, becomes pronounced 
once $R$ exceeds approximately 4 {\AA}, and approaches a negative constant at the dissociation limit.

\subsubsection{LiH dissociation} 

\begin{figure} 
\includegraphics[scale=1.12]{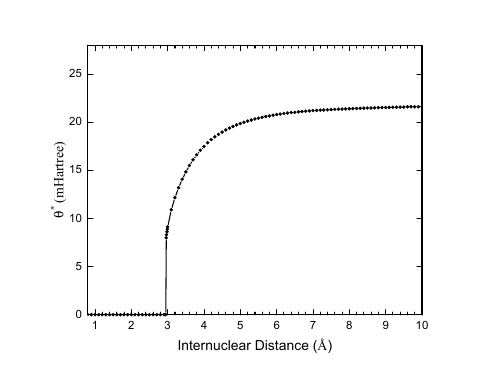} 
\caption{\label{fgr:lih_theta} 
Optimal $\theta^{*}$ values for the lowest singlet state of LiH as a function of the internuclear distance $R$, calculated using TAO-LDA within the present scheme.} 
\end{figure}

We further investigate the dissociation of LiH, a heteronuclear diatomic molecule with a single bond. First, we determine the optimal $\theta^{*}$ values for the lowest singlet state of LiH as a function of the internuclear 
distance $R$, computed using TAO-LDA (see \Cref{fgr:lih_theta}). Near the equilibrium bond length (the SR region), $\theta^{*}$ remains zero. When $R$ exceeds approximately 2.96 {\AA} (the CF point of KS-LDA), 
$\theta^{*}$ rises rapidly, indicating a transition from SR to MR character. At the dissociation limit (the MR region), $\theta^{*}$ reaches approximately 22 mHartree. 

\begin{figure} 
\includegraphics[scale=1.12]{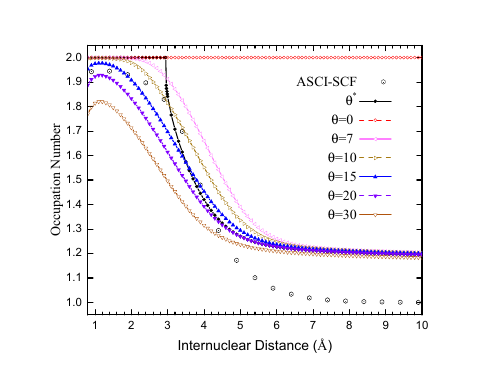} 
\caption{\label{fgr:lih_homo} 
Occupation numbers of the 2$\sigma$ orbital for the lowest singlet state of LiH as a function of the internuclear distance $R$, calculated using spin-restricted TAO-LDA with the optimal 
$\theta^{*}$ values (see \Cref{fgr:lih_theta}) and various fixed $\theta$ values (in mHartree). The $\theta = 0$ case corresponds to spin-restricted KS-LDA. 
The reference data are the NOONs of ASCI-SCF (obtained in this work using the method of Ref.\ \cite{asci}).} 
\end{figure}

To examine whether the present scheme properly captures the evolving MR character during bond dissociation, we plot the 2$\sigma$ (HOMO) occupation numbers obtained from spin-restricted TAO-LDA with various 
$\theta$ values (see \Cref{fgr:lih_homo}). The reference data comprise the NOONs from ASCI-SCF \cite{asci} calculations, employing an active space of 4 electrons in 40 orbitals (expanding the wavefunction across 
608,400 determinants). Near the equilibrium bond length, the 2$\sigma$ occupation numbers of spin-restricted TAO-LDA (with $\theta^{*}$) remain at 2.0. Once $R$ increases beyond approximately 2.96 {\AA}, the 
occupation numbers decrease rapidly, correctly reflecting the transition from SR to MR character. However, the 2$\sigma$ occupation numbers approach 1.2 (rather than 1.0, as in ASCI-SCF) at the dissociation limit, 
because the TOONs—which are strictly governed by the FD distribution function—may not perfectly match the exact NOONs in certain cases (e.g., the dissociation of a heteronuclear diatomic molecule \cite{LiH_TAO}). 
Overall, the 2$\sigma$ occupation numbers of spin-restricted TAO-LDA (with $\theta^{*}$) agree reasonably well with the ASCI-SCF NOONs, except in the region of $R > 5$ {\AA}. 

\begin{figure} 
\includegraphics[scale=1.12]{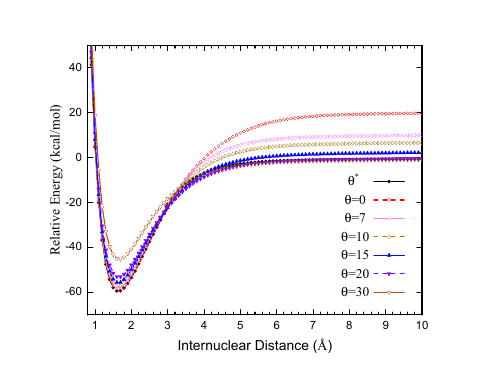} 
\caption{\label{fgr:lih_relative} 
Potential energy curves (in relative energy) for the lowest singlet state of LiH as a function of the internuclear distance $R$, calculated using spin-restricted TAO-LDA with the optimal $\theta^{*}$ 
values (see \Cref{fgr:lih_theta}) and various fixed $\theta$ values (in mHartree). The $\theta = 0$ case corresponds to spin-restricted KS-LDA. 
The zeros of energy are set at the respective spin-unrestricted dissociation limits.} 
\end{figure}

To examine how this TOON discrepancy affects the energetics, we plot the potential energy curves (in relative energy) for the lowest singlet state of LiH as a function of $R$, computed using spin-restricted TAO-LDA 
with various $\theta$ values, where the zeros of energy are set at the respective spin-unrestricted dissociation limits. As shown in \Cref{fgr:lih_relative}, similar to the previously examined systems, TAO-LDA (with 
$\theta^{*}$) performs very similarly to KS-LDA ($\theta = 0$) near the equilibrium bond length, and yields a vanishingly small SCE at the dissociation limit, regardless of the TOON discrepancy in this regime. 

\begin{figure} 
\includegraphics[scale=1.12]{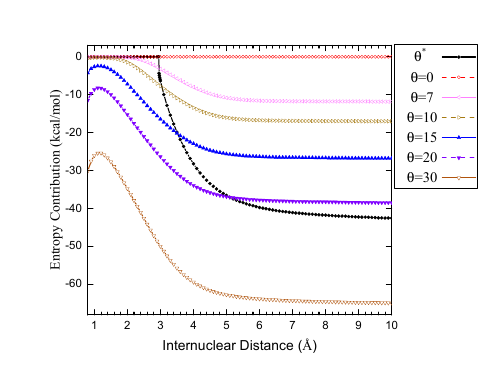} 
\caption{\label{fgr:lih_entropy} 
Entropy contributions (approximations to the static correlation energy) for the lowest singlet state of LiH as a function of the internuclear distance $R$, calculated using spin-restricted TAO-LDA 
with the optimal $\theta^{*}$ values (see \Cref{fgr:lih_theta}) and various fixed $\theta$ values (in mHartree). The $\theta = 0$ case corresponds to spin-restricted KS-LDA.} 
\end{figure}

\Cref{fgr:lih_entropy} illustrates the entropy contributions (approximations to the static correlation energy) for the lowest singlet state of LiH as a function of $R$, computed using spin-restricted TAO-LDA with various 
$\theta$ values. Consistent with the preceding systems, the entropy contribution of spin-restricted TAO-LDA (with $\theta^{*}$) is negligible near the equilibrium bond length, becomes pronounced once $R$ exceeds 
approximately 2.96 {\AA}, and approaches a negative constant at the dissociation limit, properly simulating the static correlation energy throughout bond dissociation.

\subsubsection{Twisted ethylene} 

At an HCCH torsion angle of $90^\circ$, the $\pi(1b_{2})$ and $\pi^{*}(2b_{2})$ orbitals of ethylene (C$_{2}$H$_{4}$) become degenerate, causing twisted ethylene to exhibit strong static correlation. Consequently, 
the torsion potential energy curve is known to display an unphysical cusp at the HCCH torsion angle of $90^\circ$ in spin-restricted KS-DFT calculations \cite{chai_density_2012}. We therefore also apply the present 
scheme to this system. In our calculations, we adopt the experimental geometry of C$_{2}$H$_{4}$, with $R_{\text{CC}} = 1.339$ {\AA}, $R_{\text{CH}} = 1.086$ {\AA}, and 
$\angle_\text{HCH} = 117.6^\circ$ \cite{herzberg1966}. 

\begin{figure} 
\includegraphics[scale=1.12]{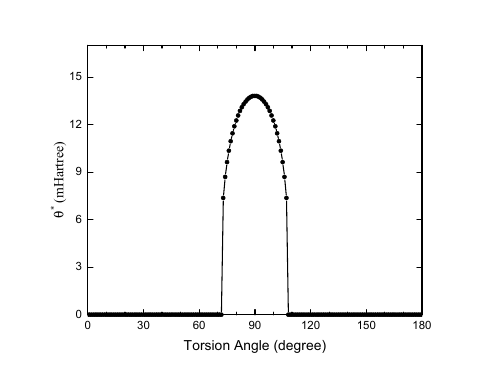} 
\caption{\label{fgr:c2h4_theta} 
Optimal $\theta^{*}$ values for the lowest singlet state of twisted ethylene as a function of the HCCH torsion angle, calculated using TAO-LDA within the present scheme.} 
\end{figure}

Using TAO-LDA within the present scheme, we determine the optimal $\theta^{*}$ values for the lowest singlet state of twisted ethylene as a function of the HCCH torsion angle (\Cref{fgr:c2h4_theta}). At small torsion 
angles, $\theta^{*}$ remains zero before rising sharply at approximately $73^\circ$, and reaching a maximum of 13.81 mHartree at $90^\circ$. As the torsion angle increases further, $\theta^{*}$ decreases 
symmetrically, returning to zero across the range from $107^\circ$ to $180^\circ$. 

\begin{figure} 
\includegraphics[scale=1.12]{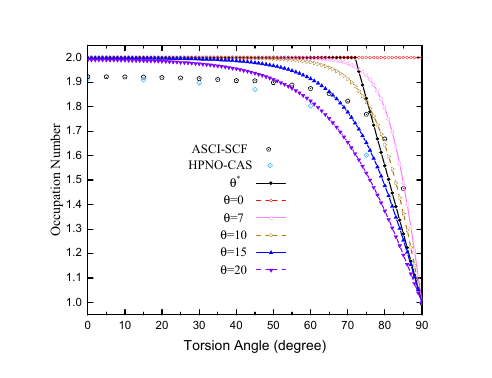} 
\caption{\label{fgr:c2h4_homo} 
Occupation numbers of the $\pi(1b_{2})$ orbital for the lowest singlet state of twisted ethylene as a function of the HCCH torsion angle, calculated using spin-restricted TAO-LDA with the optimal 
$\theta^{*}$ values (see \Cref{fgr:c2h4_theta}) and various fixed $\theta$ values (in mHartree). The $\theta = 0$ case corresponds to spin-restricted KS-LDA. 
The reference data are the HPNO-CAS (taken from Ref.\ \cite{c2h4half}) and the NOONs of ASCI-SCF (obtained in this work using the method of Ref.\ \cite{asci}).} 
\end{figure}

To examine whether the present scheme properly captures the evolving MR character upon twisting, \Cref{fgr:c2h4_homo} plots the $\pi(1b_{2})$ occupation numbers for the lowest singlet state of twisted ethylene as a 
function of the HCCH torsion angle, calculated using spin-restricted TAO-LDA with various $\theta$ values. The half-projected NOONs from CASSCF (HPNO-CAS) \cite{c2h4half} and NOONs from ASCI-SCF \cite{asci} 
serve as benchmark references; here, the ASCI-SCF calculations employ spin purification with an active space of 12 electrons in 15 orbitals, expanding the wavefunction across 2 million determinants. The $\pi(1b_{2})$ 
occupation numbers of spin-restricted TAO-LDA (with $\theta^{*}$) remain at 2.0 at small torsion angles, decrease rapidly near $73^\circ$, and approach 1.0 at $90^\circ$. Overall, the $\pi(1b_{2})$ occupation numbers 
of spin-restricted TAO-LDA (with $\theta^{*}$) are in reasonably good agreement with the HPNO-CAS and ASCI-SCF NOONs. 

\begin{figure} 
\includegraphics[scale=1.12]{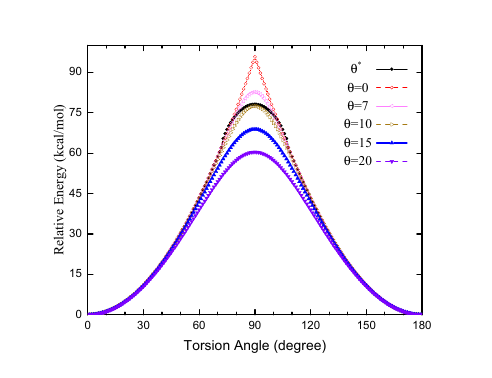} 
\caption{\label{fgr:c2h4_relative} 
Torsion potential energy curves (in relative energy) for the lowest singlet state of twisted ethylene as a function of the HCCH torsion angle, calculated using spin-restricted TAO-LDA 
with the optimal $\theta^{*}$ values (see \Cref{fgr:c2h4_theta}) and various fixed $\theta$ values (in mHartree). The $\theta = 0$ case corresponds to spin-restricted KS-LDA. 
The zeros of energy are set at the respective minimum energies.} 
\end{figure}

We further examine the torsion potential energy curves (in relative energy) for the lowest singlet state of twisted ethylene as a function of the HCCH torsion angle, calculated using spin-restricted TAO-LDA with various 
$\theta$ values, where the zeros of energy are set at the respective minimum energies. As shown in \Cref{fgr:c2h4_relative}, the KS-LDA ($\theta = 0$) curve exhibits an unphysical cusp at $90^\circ$. By contrast, 
TAO-LDA (with $\theta^{*}$)—which performs very similarly to KS-LDA ($\theta = 0$) in the SR region (torsion angles from $0^\circ$ to $73^\circ$ and $107^\circ$ to $180^\circ$)—successfully removes the unphysical 
cusp at $90^\circ$. 

\begin{figure} 
\includegraphics[scale=1.12]{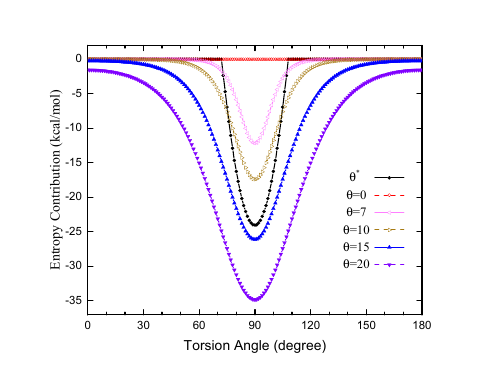} 
\caption{\label{fgr:c2h4_entropy} 
Entropy contributions (approximations to the static correlation energy) for the lowest singlet state of twisted ethylene as a function of the HCCH torsion angle, calculated using spin-restricted TAO-LDA 
with the optimal $\theta^{*}$ values (see \Cref{fgr:c2h4_theta}) and various fixed $\theta$ values (in mHartree). The $\theta = 0$ case corresponds to spin-restricted KS-LDA.} 
\end{figure}

\Cref{fgr:c2h4_entropy} illustrates the entropy contributions (approximations to the static correlation energy) for the lowest singlet state of twisted ethylene as a function of the HCCH torsion angle, calculated using 
spin-restricted TAO-LDA with various $\theta$ values. The entropy contribution of spin-restricted TAO-LDA (with $\theta^{*}$) is negligible in the SR region, and becomes pronounced near $90^\circ$.

\subsection{SR systems} 

\subsubsection{He$_{2}$ dissociation} 

The helium dimer (He$_{2}$) does not form a covalent bond and is instead weakly bound by van der Waals (vdW) interactions. Because each He atom retains its closed-shell electronic configuration, He$_{2}$ is 
expected to exhibit pure SR character across all internuclear distances. 

\begin{figure} 
\includegraphics[scale=1.12]{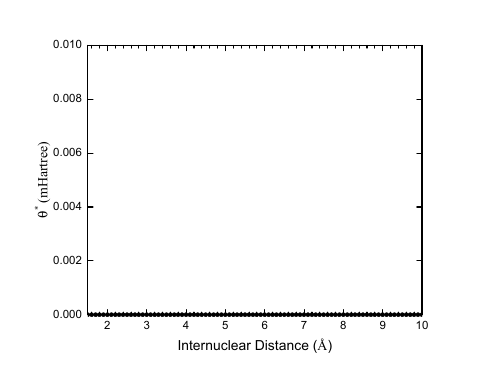} 
\caption{\label{fgr:he2_theta} 
Optimal $\theta^{*}$ values for the lowest singlet state of He$_{2}$ as a function of the internuclear distance $R$, calculated using TAO-LDA within the present scheme.} 
\end{figure}

Using TAO-LDA within the present scheme, we determine the optimal $\theta^{*}$ values for the lowest singlet state of He$_{2}$ as a function of the internuclear distance $R$ (see \Cref{fgr:he2_theta}). 
As expected, $\theta^{*} = 0$ across the entire range of $R$. 

\begin{figure} 
\includegraphics[scale=1.12]{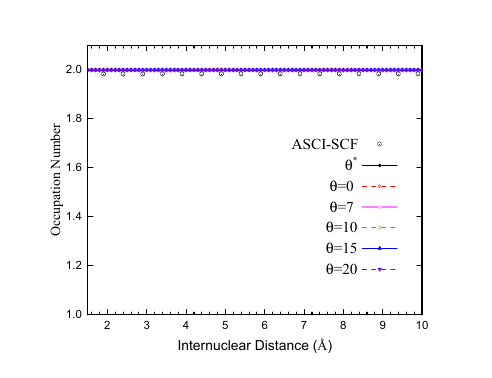} 
\caption{\label{fgr:he2_homo} 
Occupation numbers of the 1$\sigma^{*}_{u}$ orbital for the lowest singlet state of He$_{2}$ as a function of the internuclear distance $R$, calculated using spin-restricted TAO-LDA with the optimal 
$\theta^{*}$ values (see \Cref{fgr:he2_theta}) and various fixed $\theta$ values (in mHartree). The $\theta = 0$ case corresponds to spin-restricted KS-LDA. 
The reference data are the NOONs of ASCI-SCF (obtained in this work using the method of Ref.\ \cite{asci}).} 
\end{figure}

Next, we examine the occupation numbers of the 1$\sigma^{*}_{u}$ orbital (HOMO) for the lowest singlet state of He$_{2}$ across varying $R$, computed via spin-restricted TAO-LDA with various $\theta$ values 
(see \Cref{fgr:he2_homo}). For reference NOONs, the benchmark ASCI-SCF calculations employ an active space of 4 electrons in 40 orbitals, expanding the wavefunction across 608,400 determinants. 
Overall, the 1$\sigma^{*}_{u}$ occupation numbers of spin-restricted TAO-LDA (with $\theta^{*}$), which equal 2.0 across all internuclear distances, are in reasonably good agreement with the ASCI-SCF NOONs. 

\begin{figure} 
\includegraphics[scale=1.12]{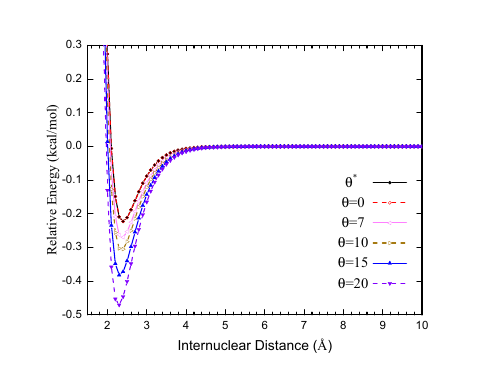} 
\caption{\label{fgr:he2_relative} 
Potential energy curves (in relative energy) for the lowest singlet state of He$_{2}$ as a function of the internuclear distance $R$, calculated using spin-restricted TAO-LDA with the optimal $\theta^{*}$ 
values (see \Cref{fgr:he2_theta}) and various fixed $\theta$ values (in mHartree). The $\theta = 0$ case corresponds to spin-restricted KS-LDA. 
The zeros of energy are set at the respective spin-unrestricted dissociation limits.} 
\end{figure}

We further examine the potential energy curves (in relative energy) for the lowest singlet state of He$_{2}$ as a function of $R$, computed using spin-restricted TAO-LDA with various $\theta$ values, where the zeros 
of energy are set at the respective spin-unrestricted dissociation limits. As shown in \Cref{fgr:he2_relative}, the curves of spin-restricted TAO-LDA (with $\theta^{*}$) and spin-restricted KS-LDA ($\theta = 0$) 
completely overlap across all internuclear distances, consistent with the pure SR character of this system. 

\begin{figure} 
\includegraphics[scale=1.12]{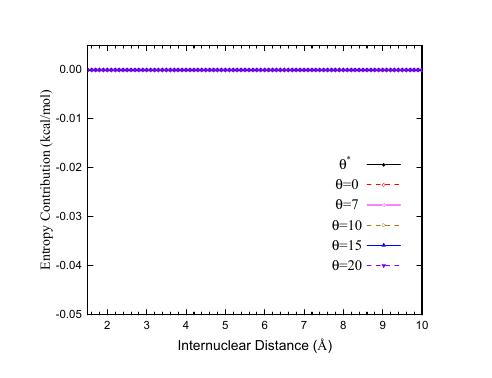} 
\caption{\label{fgr:he2_entropy} 
Entropy contributions (approximations to the static correlation energy) for the lowest singlet state of He$_{2}$ as a function of the internuclear distance $R$, calculated using spin-restricted TAO-LDA 
with the optimal $\theta^{*}$ values (see \Cref{fgr:he2_theta}) and various fixed $\theta$ values (in mHartree). The $\theta = 0$ case corresponds to spin-restricted KS-LDA.} 
\end{figure}

Finally, \Cref{fgr:he2_entropy} illustrates the entropy contributions (approximations to the static correlation energy) for the lowest singlet state of He$_{2}$ as a function of $R$, computed using spin-restricted TAO-LDA 
with various $\theta$ values. As in spin-restricted KS-LDA ($\theta = 0$), the entropy contribution of spin-restricted TAO-LDA (with $\theta^{*}$) is zero across the entire range of $R$.

\subsubsection{Ne$_{2}$ dissociation} 

\begin{figure} 
\includegraphics[scale=1.12]{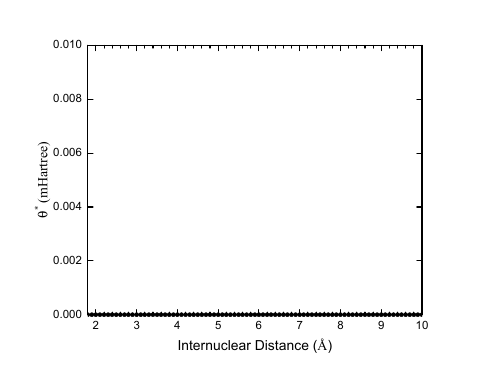} 
\caption{\label{fgr:ne2_theta} 
Optimal $\theta^{*}$ values for the lowest singlet state of Ne$_{2}$ as a function of the internuclear distance $R$, calculated using TAO-LDA within the present scheme.} 
\end{figure}

We next consider the neon dimer (Ne$_{2}$), another closed-shell noble-gas dimer expected to exhibit pure SR character. Using TAO-LDA within the present scheme, we determine the optimal $\theta^{*}$ values 
for the lowest singlet state of Ne$_{2}$ as a function of the internuclear distance $R$ (\Cref{fgr:ne2_theta}). As expected, $\theta^{*} = 0$ across the entire range of $R$. 

\begin{figure} 
\includegraphics[scale=1.12]{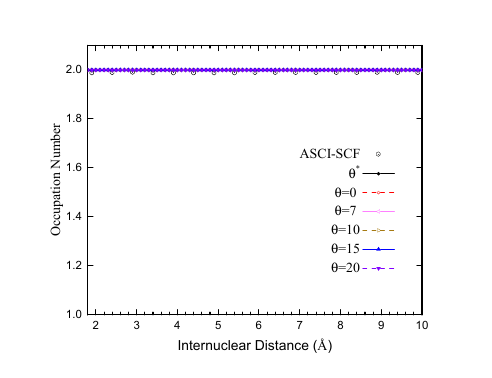} 
\caption{\label{fgr:ne2_homo} 
Occupation numbers of the 3$\sigma^{*}_{u}$ orbital for the lowest singlet state of Ne$_{2}$ as a function of the internuclear distance $R$, calculated using spin-restricted TAO-LDA with the optimal 
$\theta^{*}$ values (see \Cref{fgr:ne2_theta}) and various fixed $\theta$ values (in mHartree). The $\theta = 0$ case corresponds to spin-restricted KS-LDA. 
The reference data are the NOONs of ASCI-SCF (obtained in this work using the method of Ref.\ \cite{asci}).} 
\end{figure}

Next, we examine the occupation numbers of the 3$\sigma^{*}_{u}$ orbital (HOMO) across varying $R$, computed via spin-restricted TAO-LDA alongside reference NOONs from ASCI-SCF \cite{asci} calculations. 
Here, the ASCI-SCF calculations employ an active space of 16 electrons in 13 orbitals, expanding the wavefunction across 1,656,369 determinants. As shown in \Cref{fgr:ne2_homo}, the 3$\sigma^{*}_{u}$ occupation 
numbers of spin-restricted TAO-LDA (with $\theta^{*}$), which equal 2.0 across all internuclear distances, are in reasonably good agreement with the ASCI-SCF NOONs. 

\begin{figure} 
\includegraphics[scale=1.12]{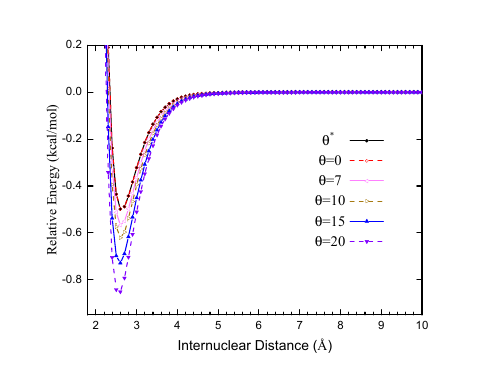} 
\caption{\label{fgr:ne2_relative} 
Potential energy curves (in relative energy) for the lowest singlet state of Ne$_{2}$ as a function of the internuclear distance $R$, calculated using spin-restricted TAO-LDA with the optimal $\theta^{*}$ 
values (see \Cref{fgr:ne2_theta}) and various fixed $\theta$ values (in mHartree). The $\theta = 0$ case corresponds to spin-restricted KS-LDA. 
The zeros of energy are set at the respective spin-unrestricted dissociation limits.} 
\end{figure}

We further examine the potential energy curves (in relative energy) for the lowest singlet state of Ne$_{2}$ as a function of $R$, computed using spin-restricted TAO-LDA with various $\theta$ 
values, where the zeros of energy are set at the respective spin-unrestricted dissociation limits. As shown in \Cref{fgr:ne2_relative}, the curves of spin-restricted TAO-LDA (with $\theta^{*}$) and spin-restricted 
KS-LDA ($\theta = 0$) completely overlap across all internuclear distances, consistent with the pure SR character of this system. 

\begin{figure} 
\includegraphics[scale=1.12]{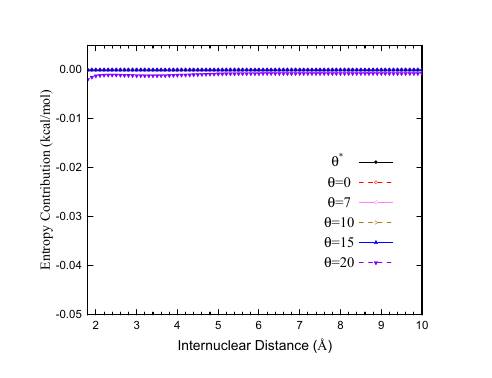} 
\caption{\label{fgr:ne2_entropy} 
Entropy contributions (approximations to the static correlation energy) for the lowest singlet state of Ne$_{2}$ as a function of the internuclear distance $R$, calculated using spin-restricted TAO-LDA 
with the optimal $\theta^{*}$ values (see \Cref{fgr:ne2_theta}) and various fixed $\theta$ values (in mHartree). The $\theta = 0$ case corresponds to spin-restricted KS-LDA.} 
\end{figure}

\Cref{fgr:ne2_entropy} illustrates the entropy contributions (approximations to the static correlation energy) for the lowest singlet state of Ne$_{2}$ as a function of $R$, computed using spin-restricted TAO-LDA with 
various $\theta$ values. As in spin-restricted KS-LDA ($\theta = 0$), the entropy contribution of spin-restricted TAO-LDA (with $\theta^{*}$) is zero across the entire range of $R$.

\section{Conclusions}\label{sec:conclusions} 

In conclusion, we have proposed a nonempirical scheme based on spin-symmetry restoration to determine the optimal fictitious temperature $\theta^{*}$ for the lowest singlet state of a molecular system within 
TAO-DFT. The resulting $\theta^{*}$ values remain vanishingly small in SR systems—where TAO-DFT naturally reduces to conventional KS-DFT—and increase appropriately in regions where MR character 
becomes prominent. 

We evaluated the performance of this scheme across both MR benchmark systems (the dissociation of H$_{2}$, N$_{2}$, Li$_{2}$, and LiH, as well as the torsion of twisted ethylene) and pure SR systems (the 
dissociation of He$_{2}$ and Ne$_{2}$). For the bond-breaking systems and twisted ethylene, $\theta^{*}$ remains negligible in the SR region and rises sharply once MR character emerges. By contrast, for the 
closed-shell noble-gas dimers (He$_{2}$ and Ne$_{2}$), $\theta^{*}$ remains zero across all internuclear distances, consistent with their pure SR character. Consequently, potential energy curves computed via 
spin-restricted TAO-LDA (with $\theta^{*}$) accurately describe the full range of molecular geometries examined. Furthermore, the TOONs obtained with $\theta^{*}$ closely track reference NOONs, while the entropy 
contributions grow systematically as static correlation strengthens. 

Overall, our approach establishes a nonempirical route to determining a system-dependent fictitious temperature without relying on empirical parameters or high-level reference data. It eliminates spurious entropy 
contributions in SR regions while introducing essential fractional orbital occupations where MR character dominates. Because the present formulation relies on spin-symmetry restoration, it is restricted to 
the lowest singlet state of a molecular system; extending this framework to establish a general criterion for arbitrary spin states remains an important direction for future work.

\section*{Author contributions} 
Conceptualization, J.-D.C.; data curation, B.-Y.L.; formal analysis, B.-Y.L.; funding acquisition, J.-D.C.; investigation, B.-Y.L.; methodology, B.-Y.L. and J.-D.C.; project administration, J.-D.C.; 
supervision, J.-D.C.; software, J.-D.C.; resources, J.-D.C.; validation, B.-Y.L. and J.-D.C.; visualization, B.-Y.L.; writing—original draft, B.-Y.L. and J.-D.C.; writing—review and editing, J.-D.C. 
All authors have read and agreed to the published version of the manuscript.

\section*{Conflicts of interest} 
There are no conflicts to declare.

\section*{Data availability} 
The data that support the findings of this study are available upon reasonable request from the authors.

\section*{Acknowledgements} 
This work was supported by National Science and Technology Council of Taiwan (Grant No.: NSTC115-2112-M-002-011), National Taiwan University, and National Center for Theoretical Sciences of Taiwan.

\section*{Statement on the use of generative AI} 
The authors acknowledge the use of Google Gemini for language polishing and readability improvements during manuscript preparation. 
The authors reviewed and edited all content and take full responsibility for the final text.

\newpage 

\end{document}